# A Frequency-domain Compensation Scheme for IQ-Imbalance in OFDM Receivers

Shu Feng, Wang Mao, Shi Xiajie, Liu Junhao, Sheng Weixing, and Xie Renhong

 Abstract—A special pilot pattern across two OFDM symbols is devised for channel estimation in OFDM systems with IQ imbalance at receiver. Based on this pilot pattern, a high-efficiency time-domain (TD) least square (LS) channel estimator is proposed to suppress channel noise by a factor of $N/(L+1)$ in comparison with the frequency-domain LS one in [1] where N and L+1 are the total number of subcarriers and the length of cyclic prefix, respectively. Following this, a low-complexity frequency-domain (FD) Gaussian elimination (GE) equalizer is proposed to eliminate IQ distortion by using only $2N$ complex multiplications per OFDM symbol. From simulation, the proposed TD-LS/FD-GE scheme using only two pilot OFDM symbols achieves the same bit error rate ( BER) performance under ideal channel knowledge and no IQ imbalances at low and medium signal-to-noise ratio (SNR) regions whereas these compensation schemes including FD-LS/Post-FFT LS, FD-LS/Pre-FFT Corr, and SPP/Pre-FFT Corr in [1] require about twenty OFDM training symbols to reach the same performance where A/B denotes compensation scheme with A being channel estimator and B being equalizer.

## I INTRODUCTION

Orthogonal frequency division multiplexing (OFDM) has been adopted in several standards such as wireless local area network (IEEE 802.11a, g and n), wireless metropolitan area network (IEEE 802.16d, e and m), digital audio broadcasting, LTE/LTE-advanced, digital radio mondiale and digital video broadcasting. Compared with the heterodyne receiver, the direct conversion RF receiving architecture is recently reconsidered as a promising solution in OFDM systems to reduce the cost and power consumption of the receiver [1]-[3]. However, the latter is severely distorted by gain and phase imbalances between the I and Q paths due to imperfections of the analog components [1]-[3]. This will

Manuscript received June 21, 2011. This work was supported in part by the open research fund of National Mobile Communications Research Laboratory (No. 2010D13), Southeast University, China.
The authors are with the Department of Communication Engineering, Nanjing University of Science and Technology, Nanjing, China. Shu Feng is also with National Mobile Communications Research Laboratory, Southeast University, China (email: shufeng@mail.njust.edu.cn, wangmao123@gmail.com).

severely destroy the orthogonality among the OFDM subcarriers and cause intercarrier interference, giving rise to a high bit error rate (BER) floor. Therefore, estimation and compensation of IQ imbalance in the direct conversion receivers are crucial to OFDM receiver performance.

The schemes of canceling IQ imbalance have been investigated by several scholars. In [1], the authors derive the SNR loss of IQ-imbalance in OFDM receivers and propose several frequency domain(FD) and time-domain (TD) methods including post-FFT least-squares, adaptive least mean square (LMS) and pre-FFT TD compensation to eliminate IQ distortions. They extend these methods to IQ imbalances at both transmitter and receiver [4]. Blind estimation and compensation schemes in the time domain have also been proposed [5]. Joint estimation of IQ imbalance and several other impairments such as phase noise, frequency offset are investigated in [6]-[10]. In [6], a finite impulse response (FIR) filter followed by an asymmetric phase compensator has been proposed to correct both frequency dependent and frequency independent IQ imbalance. In [8], a differential filter is employed to estimate the frequency offset and IQ imbalance. A compensation method based on the subcarrier allocation of OFDM signals is proposed in [9]. [11] extends the research of Tx/Rx IQ imbalances to the case of packet-switched systems. In [12] and [13], authors focus on pilot design and reduced complexity compensation in MIMO-OFDM systems with IQ imbalance.

Unfortunately, the FD LS channel estimation in [1] doesn't exploit the TD property of the channel in the presence of IQ imbalance. Thus, it requires more than twenty training OFDM symbols to achieve the same BER performance with ideal channel knowledge and no IQ-imbalance (abbreviated as ideal IQ below). Obviously, this scheme is low on bandwidth efficiency. To overcome this problem, we design an LS channel estimator which fully exploits the TD property of channel parameters to reduce the impact of channel noise. Hence, it requires only two OFDM symbols to reach the BER performance of ideal IQ.

Notations: Bold letters denote vectors and matrices. $(\cdot)^T$, $(\cdot)^*$, and $(\cdot)^H$ denotes transpose, conjugate, and conjugate transpose operations, respectively. Operation diag($\mathbf{x}$) places vector $\mathbf{x}$ on diagonal of a diagonal matrix. $\mathbf{I}_n$ and $\mathbf{0}_{n \times m}$ are the $n \times n$ identity and $n \times m$ zero matrices, respectively.

This paper is organized as follows. Section II describes the system model. Pilot pattern, TD-LS channel estimation and Gaussian elimination (GE) equalization are proposed in section III. Simulation results are listed in section IV. Section V concludes the paper.

## II SYSTEM MODEL

In OFDM systems with IQ imbalance as shown in Fig.1, the transmitted block of $N$ data symbols over $N$ subcarriers is denoted as

$$\mathbf{s} = \left[ s(1)\, s(2) \cdots s(N) \right]^T \tag{1}$$

whose IDFT operation yields

$$\overline{\mathbf{s}} = \mathbf{F}^H \mathbf{s} \tag{2}$$

where

$$\mathbf{F}(m,n) = \frac{1}{\sqrt{N}} \exp\left( \frac{-j 2\pi (n-1)(m-1)}{N} \right) \tag{3}$$

$$j = \sqrt{-1} \qquad m,n \in \{1,2,\cdots,N\}$$

Then, the received OFDM symbol before being distorted by IQ imbalance is expressed as

$$\overline{\mathbf{y}} = \mathbf{F}^H \mathbf{\Lambda} \mathbf{F}\, \overline{\mathbf{s}} + \overline{\mathbf{w}} \tag{4}$$

where

$$\mathbf{\Lambda} = \mathrm{diag}\{\mathbf{H}\} \ \text{ and } \ \mathbf{H} = \mathbf{F}\begin{pmatrix} \mathbf{h} \\ \mathbf{0}_{(N-L-1)\times 1} \end{pmatrix} \tag{5}$$

and $\mathbf{h} = [h(1)\, h(2)\cdots h(L+1)]^T$ is the channel impulse response (CIR). The received OFDM symbol distorted by IQ imbalance is written as

$$\overline{\mathbf{z}} = \mu\overline{\mathbf{y}} + v\overline{\mathbf{y}}^* \tag{6}$$

where $\mu = \cos(\theta/2) + j\alpha\sin(\theta/2)$ and $v = \alpha\cos(\theta/2) - j\sin(\theta/2)$, $\theta$ and $\alpha$ are phase and amplitude imbalance between I and Q branches[1].Taking FFT operation on (6) gives

$$\mathbf{z} = \mu\,\mathrm{diag}\{\mathbf{H}\}\mathbf{s} + v\,\mathrm{diag}\{\mathbf{H}^{\#}\}\mathbf{s}^{\#} + \mathbf{w} \tag{7}$$

where the operation # is defined as [1]

$$\mathbf{X}^{\#} = [X^*(1) \ X^*(2) \cdots X^*(N/2+2) \ X^*(N/2+1) \ X^*(N/2) \cdots X^*(2)]^{\mathrm{T}} \tag{8}$$

where

$$\mathbf{X} = [X(1) \ X(2) \cdots X(N/2) \ X(N/2+1) \ X(N/2+2) \cdots X(N)]^{\mathrm{T}} \tag{9}$$

From [1], if $\mathbf{X} = \mathbf{F}\mathbf{x}$, then

$$\mathbf{X}^{\#} = (\mathbf{F}\mathbf{x})^{\#} = \mathbf{F}\mathbf{x}^* \tag{10}$$

Thus, we obtain the following equality

$$\left(\mathbf{X}^{\#}\right)^{\#} = \left(\mathbf{F}\mathbf{x}^*\right)^{\#} = \mathbf{F}\mathbf{x}^{**} = \mathbf{X} \tag{11}$$

For the simplicity of discussion, block fading is assumed in the following, i.e., channel is assumed to be constant within a frame and variable from frame to frame.

## III PROPOSED SCHEME COMBINING PILOT DESIGN, EQUALIZATION AND CHANNEL ESTIMATION

In the following, a low-complexity Gaussian elimination is adopted to cancel the IQ distortion based on operation #. Then, a particular training pattern using two adjacent OFDM symbols is designed and a TD-LS channel estimation is presented to provide a high-precision estimation of channel parameters $v/\mu^*$, $\mu \mathbf{H}$ and $v^* \mathbf{H}$.

### A. Gaussian Elimination Equalizer

Due to the equalities, $\left(\mathrm{diag}\{\mathbf{H}\}\mathbf{s}\right)^{\#} = \mathrm{diag}\{\mathbf{H}^{\#}\}\mathbf{s}^{\#}$ and $\left(\mathrm{diag}\{\mathbf{H}^{\#}\}\mathbf{s}^{\#}\right)^{\#} = \mathrm{diag}\{\mathbf{H}\}\mathbf{s}$, making # operation on (7) yields

$$\mathbf{z}^{\#} = v^* \mathrm{diag}\{\mathbf{H}\}\mathbf{s} + \mu^* \mathrm{diag}\{\mathbf{H}\}^{\#}\mathbf{s}^{\#} + \mathbf{w}^{\#} \tag{12}$$

Assuming $\kappa = \dfrac{v}{\mu^*}$ is known, based on (7) and (12), we construct

$$\mathbf{z} - \kappa \mathbf{z}^{\#} = \left(\mu - \kappa v^*\right)\mathrm{diag}\{\mathbf{H}\}\mathbf{s} + \mathbf{w} - \kappa \mathbf{w}^{\#} \tag{13}$$

(13) no longer has the IQ distortion present in (7) and can be rewritten as

$$\mathbf{z} - \kappa \mathbf{z}^{\#} = \left(\mathrm{diag}\{\mu \mathbf{H}\} - \kappa \mathrm{diag}\{v^* \mathbf{H}\}\right)\mathbf{s} + \mathbf{w} - \kappa \mathbf{w}^{\#} \tag{14}$$

which gives the following detector as

$$\hat{\mathbf{s}} = \left\{ \mathrm{diag}\left\{\mu\mathbf{H}\right\} - \kappa \mathrm{diag}\left\{v^*\mathbf{H}\right\} \right\}^{-1} \left(\mathbf{z} - \kappa\mathbf{z}^{\#}\right) \tag{15}$$

Equation (15) can be simplified as

$$\hat{\mathbf{s}}(k) = \frac{\mathbf{z}(k) - \kappa\mathbf{z}^{\#}(k)}{\mu\mathbf{H}(k) - \kappa v^*\mathbf{H}(k)} \tag{16}$$

To obtain $\mathbf{s}$ from (15) or (16), $\kappa$, $\mu\mathbf{H}$, and $v^*\mathbf{H}$ need to be estimated in advance, among which $\mu\mathbf{h}$ and $v^*\mathbf{h}$ are

$$\mu\mathbf{H} = \mathbf{F}\begin{pmatrix} \mu\mathbf{h} \\ \mathbf{0}_{(N-L-1)\times 1} \end{pmatrix} \tag{17}$$

and

$$v^*\mathbf{H} = \mathbf{F}\begin{pmatrix} v^*\mathbf{h} \\ \mathbf{0}_{(N-L-1)\times 1} \end{pmatrix} \tag{18}$$

Similar to [1], the loss in signal-to-noise (SNR) from the difference between the error variance given by (16) and the error variance $\sigma_w^2/\|\mathbf{H}(k)\|^2$ is

$$\mathrm{Loss\ in\ SNR} = 10\log\left(\frac{1 + \|\kappa\|^2}{\|\mu\|^2 - 2\mathrm{Re}\left(\kappa v^*\mu\right) + \|\kappa\|^2\|v\|^2}\right) \tag{19}$$

where Re ($\mathbf{x}$) denotes the real part of $\mathbf{x}$.

Fig. 2 shows the theoretical lower bounds concerning the loss in SNR due to IQ imbalances. The 2D surfaces of loss in SNR are based on the results (19) derived in Section III and (31) in [1]. These bounds are computed with perfect channel and distortion parameter knowledge available at the receiver, therefore serving as the theoretical lower bounds on the SNR loss due to imbalances. From this figure, the lower bound of the GE equalizer proposed in this letter is better than that of the LS equalizer in [1].

### B. Pilot Pattern Design and TD-LS Estimation of Channel Parameters.

Let us devise the frequency-domain pilot vectors of two pilot OFDM symbols at the beginning part of frame as

$$\mathbf{s_1} = \begin{pmatrix} \eta \\ \mathbf{s_p} \\ \eta \\ \mathbf{0}_{(N/2\text{-}1)\times1} \end{pmatrix} \tag{20}$$

and

$$\mathbf{s_2} = \begin{pmatrix} j\eta \\ \mathbf{0}_{(N/2\text{-}1)\times1} \\ j\eta \\ \mathbf{s_p} \end{pmatrix} \tag{21}$$

where $\eta = 2\bar{P}_s$ with $\bar{P}_s$ being the average transmit power for signal constellation and $\mathbf{s_p}$ is an $N/2-1$ dimensional column pilot vector with $\text{tr}\{\text{E}(\mathbf{s_p s_p}^H)\} = (N-1)\bar{P}_s$. After the two pilot symbol vectors passes through multipath channel, we get the following received training vectors and symbols in frequency domain as follows

$$\mathbf{z_1}(2:N/2) = \mu\text{diag}\{\mathbf{H}(2:N/2)\}\mathbf{s_p} + \mathbf{w_1}(2:N/2) \tag{22}$$

$$\mathbf{z_1}(N/2+2:N) = v\text{diag}\{\mathbf{H}^{\#}(N/2+2:N)\}\dot{\mathbf{s}}_p + \mathbf{w_1}(N/2+2:N) \tag{23}$$

$$\mathbf{z_2}(2:N/2) = v\text{diag}\{\mathbf{H}^{\#}(2:N/2)\}\dot{\mathbf{s}}_p + \mathbf{w_2}(2:N/2) \tag{24}$$

$$\mathbf{z_2}(N/2+2:N) = \mu\text{diag}\{\mathbf{H}(N/2+2:N)\}\mathbf{s_p} + \mathbf{w_2}(N/2+2:N) \tag{25}$$

$$\text{and} \quad \mathbf{z_1}(1) = \mu\mathbf{H}(1)\eta + v\mathbf{H}(1)^*\eta + \mathbf{w_1}(1) \tag{26}$$

$$\mathbf{z_1}(N/2+1) = \mu\mathbf{H}(N/2+1)\eta + v\mathbf{H}(N/2+1)^*\eta + \mathbf{w_1}(N/2+1) \tag{27}$$

$$\mathbf{z_2}(1) = j\mu\mathbf{H}(1)\eta - jv\mathbf{H}(1)^*\eta + \mathbf{w_2}(1) \tag{28}$$

$$\mathbf{z_2}(N/2+1) = j\mu\mathbf{H}(N/2+1)\eta - jv\mathbf{H}(N/2+1)^*\eta + \mathbf{w_2}(N/2+1) \tag{29}$$

Then, combining (26)-(29) forms the following equations

$$-0.5j\mathbf{z_2}(1) + 0.5\mathbf{z_1}(1) = \mu\mathbf{H}(1)\eta - 0.5j\mathbf{w_2}(1) + 0.5\mathbf{w_1}(1) \tag{30}$$

$$0.5j\mathbf{z_2}(1) + 0.5\mathbf{z_1}(1) = v\mathbf{H}(1)^*\eta + 0.5j\mathbf{w_2}(1) + 0.5\mathbf{w_1}(1) \tag{31}$$

$$-0.5j\mathbf{z_2}(N/2+1) + 0.5\mathbf{z_1}(N/2+1) = \mu\mathbf{H}(N/2+1)\eta - 0.5j\mathbf{w_2}(N/2+1) + 0.5\mathbf{w_1}(N/2+1) \tag{32}$$

$$0.5j\mathbf{z_2}(N/2+1) + 0.5\mathbf{z_1}(N/2+1) = v\mathbf{H}(N/2+1)^*\eta + 0.5j\mathbf{w_2}(N/2+1) + 0.5\mathbf{w_1}(N/2+1) \tag{33}$$

where $\dot{\mathbf{s}}_{\mathbf{p}} = \mathbf{s}_1^{\#}(N/2+2:N)$. Stacking (22), (25), (30) and (32) gives a large matrix-vector form of

$$\tilde{\mathbf{z}}_{\mathbf{a}} = \begin{pmatrix} -0.5j\mathbf{z}_2(1)+0.5\mathbf{z}_1(1) \\ \mathbf{z}_1(2:N/2) \\ -0.5j\mathbf{z}_2(N/2+1)+0.5\mathbf{z}_1(N/2+1) \\ \mathbf{z}_2(N/2+2:N) \end{pmatrix} =$$

$$\operatorname{diag}\{\mu\mathbf{H}\}\begin{pmatrix} \eta \\ \mathbf{s}_{\mathbf{p}} \\ \eta \\ \mathbf{s}_{\mathbf{p}} \end{pmatrix} + \underbrace{\begin{pmatrix} -0.5j\mathbf{w}_2(1)+0.5\mathbf{w}_1(1) \\ \mathbf{w}_1(2:N/2) \\ -0.5j\mathbf{w}_2(N/2+1)+0.5\mathbf{w}_1(N/2+1) \\ \mathbf{w}_2(N/2+2:N) \end{pmatrix}}_{\tilde{\mathbf{w}}_{\mathbf{a}}} = \operatorname{diag}\underbrace{\begin{bmatrix} \eta \\ \mathbf{s}_{\mathbf{p}} \\ \eta \\ \mathbf{s}_{\mathbf{p}} \end{bmatrix}}_{\tilde{\mathbf{s}}_{\mathbf{p}}}(\mu\mathbf{H})+\tilde{\mathbf{w}}_{\mathbf{a}} = \operatorname{diag}\{\tilde{\mathbf{s}}_{\mathbf{p}}\}\mathbf{F}\begin{pmatrix} \mu\mathbf{h} \\ \mathbf{0}_{(N-L-1)\1} \end{pmatrix}+\tilde{\mathbf{w}}_{\mathbf{a}} \quad (34)$$

Thus, the LS estimate of $\mu\mathbf{h}$ is given as

$$\widehat{\mu\mathbf{h}}_{TD-LS} = \mathbf{PF}^H\operatorname{diag}\{\tilde{\mathbf{s}}_{\mathbf{p}}\}^{-1}\tilde{\mathbf{z}}_{\mathbf{a}} = \mu\mathbf{h} + \mathbf{PF}^H\operatorname{diag}\{\tilde{\mathbf{s}}_{\mathbf{p}}\}^{-1}\tilde{\mathbf{w}}_{\mathbf{a}} \quad (35)$$

where $\mathbf{P} = \begin{bmatrix} \mathbf{I}_{L+1} & \mathbf{0}_{(L+1)\times(N-L-1)} \end{bmatrix}$. Then, we have the estimate of $\mu\mathbf{H}$ as

$$\widehat{\mu\mathbf{H}}_{TD-LS} = \mathbf{F}\begin{pmatrix} \widehat{\mu\mathbf{h}}_{LS} \\ \mathbf{0}_{(N-L-1)\times 1} \end{pmatrix} \quad (36)$$

In the same manner, combining (23), (24), (31), and (33) into a large matrix-vector form yields

$$\tilde{\mathbf{z}}_b^{\#} = \begin{pmatrix} 0.5j\mathbf{z}_2(1)+0.5\mathbf{z}_1(1) \\ \mathbf{z}_2(2:N/2) \\ 0.5j\mathbf{z}_2(N/2+1)+0.5\mathbf{z}_1(N/2+1) \\ \mathbf{z}_1(N/2+2:N) \end{pmatrix} = \operatorname{diag}\{\nu\mathbf{H}^{\#}\}\underbrace{\begin{pmatrix} \eta \\ \dot{\mathbf{s}}_{\mathbf{p}} \\ \eta \\ \dot{\mathbf{s}}_{\mathbf{p}} \end{pmatrix}}_{\dot{\mathbf{s}}_{\mathbf{p}}} + \underbrace{\begin{pmatrix} 0.5j\mathbf{w}_2(1)+0.5\mathbf{w}_1(1) \\ \mathbf{w}_2(2:N/2) \\ 0.5j\mathbf{w}_2(N/2+1)+0.5\mathbf{w}_1(N/2+1) \\ \mathbf{w}_1(N/2+2:N) \end{pmatrix}}_{\tilde{\mathbf{w}}_b} \quad (37)$$

whose # operation forms

$$\tilde{\mathbf{z}}_{\mathbf{b}} = \left(\tilde{\mathbf{z}}_{\mathbf{b}}^{\#}\right)^{\#} = \operatorname{diag}\{\nu^*\mathbf{H}\}\tilde{\dot{\mathbf{s}}}_{\mathbf{p}}^{\#} + \tilde{\mathbf{w}}_b^{\#} = \operatorname{diag}\{\tilde{\mathbf{s}}_{\mathbf{p}}\}\mathbf{F}\begin{pmatrix} (\nu^*h) \\ \mathbf{0}_{(N-L-1)\times 1} \end{pmatrix}+\tilde{\mathbf{w}}_b^{\#} \quad (38)$$

which gives the LS estimate of $\nu^*\mathbf{h}$

$$\widehat{\nu^*\mathbf{h}}_{TD-LS} = \mathbf{PF}^H\operatorname{diag}\{\tilde{\mathbf{s}}_{\mathbf{p}}\}^{-1}\tilde{\mathbf{z}}_{\mathbf{b}} = \nu^*\mathbf{h} + \mathbf{PF}^H\operatorname{diag}\{\tilde{\mathbf{s}}_{\mathbf{p}}\}^{-1}\tilde{\mathbf{w}}_{\mathbf{b}}^{\#} \quad (39)$$

Then, we have the estimate of $\nu^*\mathbf{H}$ as

$$\widehat{\nu^*\mathbf{H}}_{TD-LS} = \mathbf{F}\begin{pmatrix} \widehat{\nu^*\mathbf{h}}_{TD-LS} \\ \mathbf{0}_{(N-L-1)\times 1} \end{pmatrix} \quad (40)$$

In terms of (35), (36), (39), and (40), the estimate of $\kappa$ can be formulated as

$$\hat{\kappa}_{LS} = \frac{\sum_{k=1}^{N} v^* \hat{\overset{\wedge}{\mathbf{H}}}_{TD-LS}^*(k)}{\sum_{k=1}^{N} \mu \hat{\overset{\wedge}{\mathbf{H}}}_{TD-LS}^*(k)} = \frac{\sum_{k=1}^{L+1} v^* \hat{\overset{\wedge}{\mathbf{h}}}_{TD-LS}^*(k)}{\sum_{k=1}^{L+1} \mu \hat{\overset{\wedge}{\mathbf{h}}}_{TD-LS}^*(k)} \tag{41}$$

From (35), (36), (39), and (40), we obtain the estimation mean square errors of $v^* \mathbf{H}$ and $\mu \mathbf{H}$ as follows

$$\frac{\mathrm{E}\left\{\left(\mu \hat{\mathbf{H}}_{TD-LS} - \mu \mathbf{H}\right)^H \left(\mu \hat{\mathbf{H}}_{TD-LS} - \mu \mathbf{H}\right)\right\}}{N} = \frac{\mathrm{E}\left\{\left(v^* \hat{\overset{\wedge}{\mathbf{H}}}_{TD-LS}^* - v^* \mathbf{H}\right)^H \left(v^* \hat{\overset{\wedge}{\mathbf{H}}}_{TD-LS}^* - v^* \mathbf{H}\right)\right\}}{N} = \frac{(L+1)\beta}{N\gamma} \tag{42}$$

where $\gamma$ is signal-to-noise ratio (SNR) and is defined as $\mathrm{E}\left\{\mathbf{s}(k)^* \mathbf{s}(k)\right\}/\sigma_n^2$ [14], and

$$\beta = \frac{\mathrm{E}\left\{\mathbf{s}(k)^* \mathbf{s}(k)\right\}}{\mathrm{E}\left\{\left(1/\mathbf{s}(k)\right)^* \left(1/\mathbf{s}(k)\right)\right\}} \tag{43}$$

## IV SIMULATION AND DISCUSSION

In the following, a typical OFDM system is simulated to evaluate the performance of the proposed scheme against an ideal IQ OFDM receiver, a receiver with no compensation scheme, and those aforementioned compensation schemes, FD-LS/Post-FFT LS, FD-LS/Pre-FFT Corr, and SPP/Pre-FFT Corr in [1]. where A/B denotes compensation scheme with A being channel estimator and B being equalizer. Simulation parameters were : OFDM symbol length N=128 , cyclic prefix L+1=16 , signal bandwidth BW=2MHz, digital modulation QPSK, carrier frequency $f_c$=2GHz. A typical urban (TU) channel was employed in the simulation as in [15].

Figs. 3 to 4 compare the proposed scheme with the FD-LS channel estimator plus Post-FFT LS equalizer (FD-LS/Post-FFT) in [1], for different values of IQ imbalance parameters where $N_T$ denotes the number of consecutive training OFDM symbols (TOSs) with all subcarriers carrying pilot symbols at the beginning of each frame as shown in [1]. Here, our scheme uses two TOSs as shown in (21) and (20). From these figures, it is evident that the proposed scheme with only two TOSs achieves the same BER performance as ideal IQ at low medium SNRs whereas the LS scheme in [1] costs about $N_T$ =2 TOSs to realize almost the same BER performance. Therefore, the proposed scheme is more effective in the sense of overhead.

Figs. 5 and 6 plot the curves of BER versus SNR of the proposed TD-LS/FD-GE scheme, the FD-LS/Post-FFT LS, the FD-LS channel estimator plus pre-FFT distortion correction (FD-LS/Pre-FFT Corr), and the special pilot structure based channel estimator plus pre-FFT distortion correction (SPP/Pre-FFT Corr) in [1] for different values of IQ imbalance parameters where $N_T = 2$. In these two figures, our scheme is obviously better on BER performance than the FD-LS/Post-FFT, the FD-LS/Pre-FFT Corr, and SPP/Pre-FFT Corr in [1].

The complexity of three channel estimators the proposed TD-LS, FD-LS in [1] and the SPP in [1] are

$$4(L+1)N^2 + 2(L+1)N + 4/3N^3 + N\left(\log_2 N + \log_2(L+1)\right), \quad 36NN_T + 64/3N + 0.5NN_T \log_2 N, \quad \text{and}$$

$2(N_T + 1) + 0.5NN_T \log_2 N$ complex multiplications (CMs) where $N_F$ is the total number of non-training OFDM symbols. Clearly, three channel estimation has almost complexity. The computational amounts of three equalizer FD-GE, Post-FFT LS, and Pre-FFT Corr are $3NN_F + 0.5NN_F \log_2 N$, $32/3N + 2NN_F + 0.5NN_F \log_2 N$, and $\left(2N + 0.5N\log_2 N\right)N_F$ CMs. The proposed FD-GE and Pre-FFT are slight lower on complexity than Post-FFT LS. Thus, we conclude that four schemes including the proposed TD-LS/FD-GE scheme, the FD-LS/Post-FFT LS, FD-LS/Pre-FFT Corr, and SPP/Pre-FFT Corr has almost the same computational amount. Hence, our scheme is very attractive for mitigating IQ imbalance in practical OFDM receivers.

Additionally, our scheme can be applied to the case of frequency-dependent IQ imbalance parameters $\mu$ and $\nu^*$ like the FD-LS /Post-FFT LS scheme in [1] ($\mu$ and $\nu^*$ depends on the subcarriers $k$ (frequency-domain), these time-domain can not solve this problem, in general, when bandwidth $<$20MHz, they can be viewed as constants [1]). However, these TD compensation schemes based on pre-FFT (TD) distortion correction are not suitable for this case [1].

## V CONCLUSIONS

In this paper, a compensation scheme combining a TD-LS channel estimator and a FD GE equalizer is investigated in OFDM systems with IQ-imbalance at receiver. Compared with the FD-LS/Post-FFT LS, SPP/Pre-FFT Corr, and FD-LS/Pre-FFT Corr schemes in [1], this scheme shows better BER performance. More importantly, it needs only two OFDM training symbols to achieve the same BER performance as ideal IQ in the low and medium SNR regions. The proposed TD-LS/FD-GE can function in the case of frequency-dependent distortion parameters $\mu$ and $\nu^*$. However, the schemes based on pre-FFT correction lack this capability. Due to a short training pattern, the proposed scheme can be directly applied to time-variant wireless channels.

REFERENCES

[1] A. Tarighat, R. Bagheri, and A. H. Sayed, "Compensation schemes and performance analysis of IQ imbalances in OFDM Receivers," *IEEE Trans Signal Processing,* vol.53 , no.8, pp. 3257-3268, 2005.

[2] B. Razavi, RF Microelectronics. Englewood Cliffs, NJ: Prentice-Hall, 1998.

[3] A. A. Abidi, "Direct-conversion radio transceivers for digital communications," *IEEE J. Solid-State Circuits, ,* vol. 30, no. 12, pp. 1399-1410, Dec. 1995.

[4] A. Tarighat and A. H. Sayed, "Joint compensation of transmitter and receiver impairments in OFDM systems," *IEEE Trans Wireless Communications,,* vol. 6, no. 1, pp. 240-247, Jan. 2007.

[5] M. Valkama, M. Renfors, and V. Koivunen, "Advanced methods for IQ imbalance compensation in communication receivers," *IEEE Trans Signal Processing,* vol. 49, no. 10, pp. 2335-2344, Oct. 2001.

[6] G. Xing, M. Shen, and H. Liu, "Frequency offset and IQ imbalance compensation for direct conversion receivers," *IEEE Trans Wireless Communications,* vol. 4, no. 3, pp. 673-680, Mar. 2005.

[7] J. Tubbax, B. Come, L. Van der Perre, S. Donnay, M. Engels, H. De Man, and M. Moonen, "Compensation of IQ imbalance and phase noise in OFDM systems, " *IEEE Trans Wireless Communications,* vol. 4, no.3, pp. 872 - 877, May 2005.

[8] M. Inamori, A. M. Bostamam, Y. Sanada, and H. Minami, "IQ imbalance compensation scheme in the presence of frequency offset and dynamic DC offset for a direct conversion Receiver," *IEEE Trans Wireless Communications,* vol.8 , no.5, pp. 2214-2220, May 2009.

[9] H. Lin and K. Yamashita, "Subcarrier allocation based compensation for carrier frequency offset and IQ imbalances in OFDM systems," *IEEE Trans Wireless Communications,* vol.8 , no.1, pp. 18-23, Jan. 2009.

[10] D. Tandur and M. Moonen, "Joint adaptive compensation of transmitter and receiver IQ imbalance under carrier frequency offset in OFDM-based systems,"*IEEE Trans Signal Processing,* vol. 55, no. 11, pp. 5246 - 5252 , Nov. 2007.

[11] J. Feigin and D. Brady, "Joint Transmitter/Receiver IQ Imbalance Compensation for Direct Conversion OFDM in Packet-Switched Multipath Environments," *IEEE Trans Signal Processing,* vol.57 , no.11, pp. 4588-4593, Nov. 2009.

[12] B. Narasimhan, S. Narayanan, H. Minn, and N. Al-Dhahir, "Reduced Complexity Baseband Compensation of Joint Tx/Rx IQ Imbalance in Mobile MIMO-OFDM," *IEEE Transactions on Wireless Communications*, vol. 9, no. 5, pp. 1720-1728, May 2010.

[13] H. Minn and D. Munoz,"Pilot Designs for Channel Estimation of MIMO OFDM Systems with Frequency-Dependent IQ Imbalances," *IEEE Transactions on Communications*, vol. 58, no. 8, pp. 2252-2264, Aug. 2010.

[14] O. Edfors, M. Sandell, Jan-Jaap van de Beek, and P. O. Borjesson, "OFDM Channel Estimation by Singular Value Decomposition," *IEEE Trans. Communications*, vol. 46, no. 7, pp. 931–939, Jul. 1998.

[15] ETSI TR 125 943, Universal Mobile Telecommunications System (UMTS); Deployment (3GPP TR 25.943 Version 5.1.0 Release 5), June 2002.

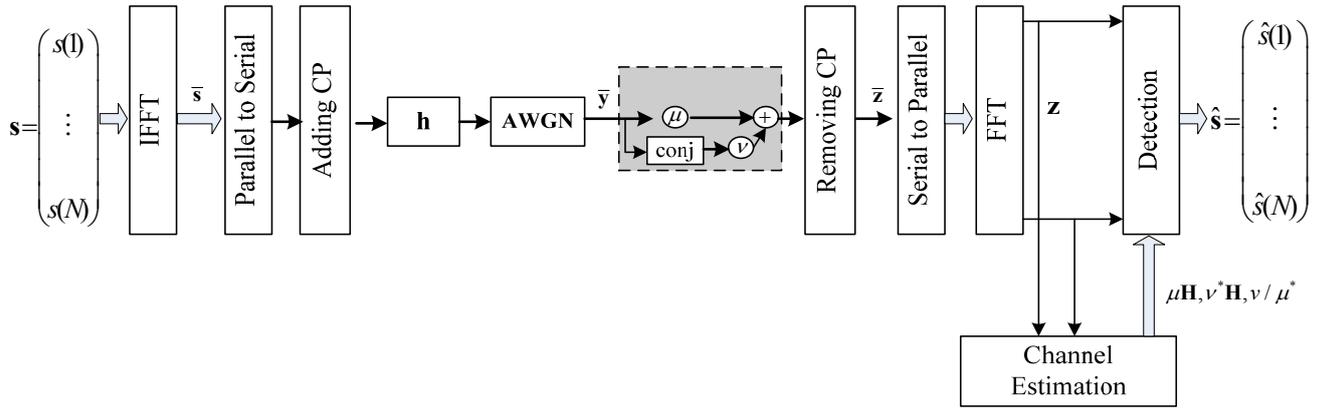

Fig. 1.Discrete baseband OFDM systems with IQ imbalance at receiver

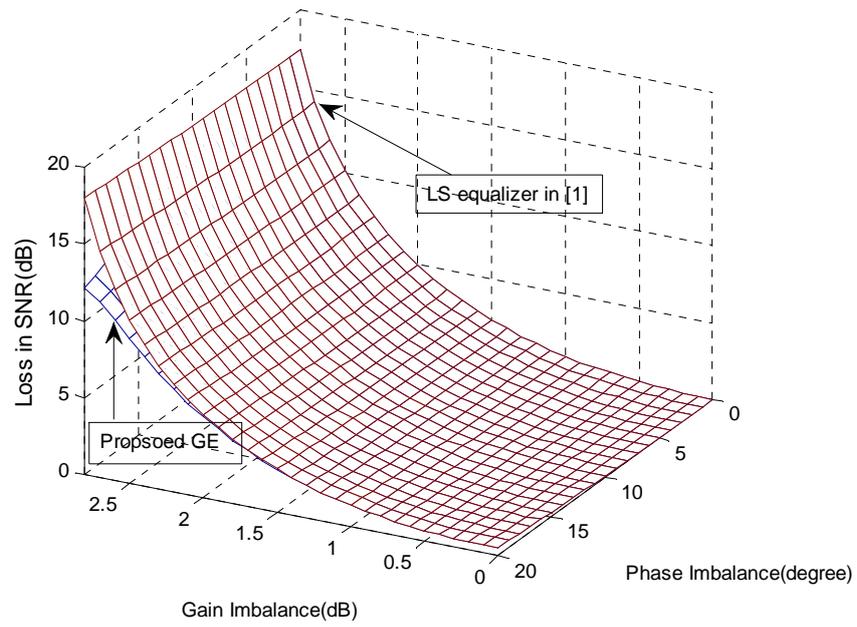

Fig. 2  Comparison of loss in SNR

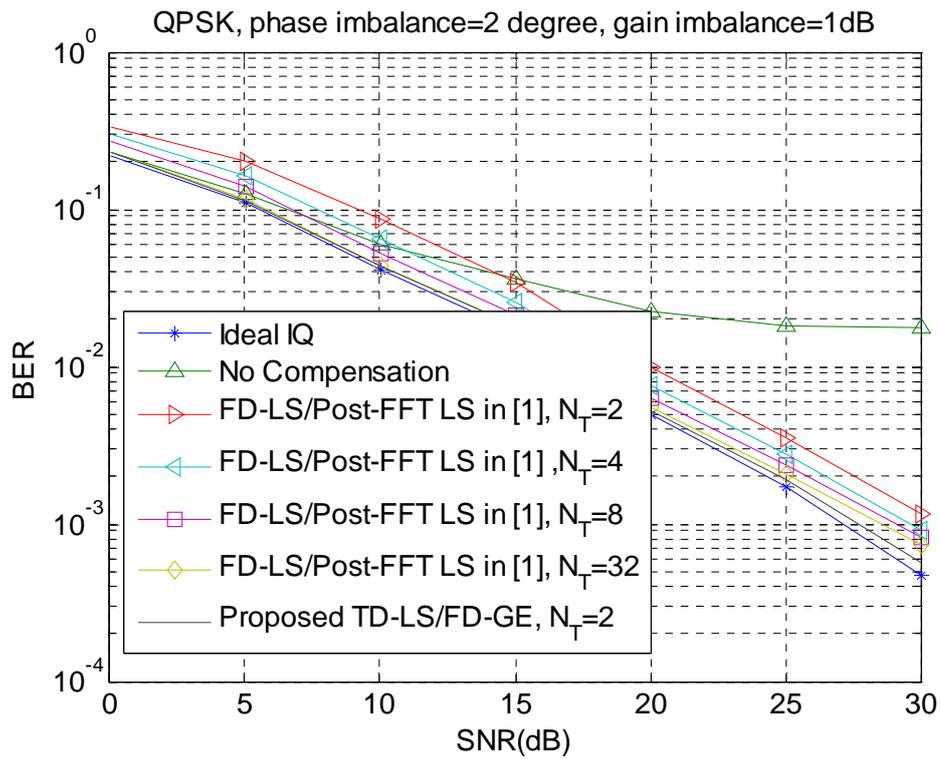

Fig. 3  Comparison of BER performance for the proposed scheme with two training OFDM

symbols and the LS in [1] with different numbers of training symbols in the case of $\theta=2^{o}$ and $\alpha=1$

dB

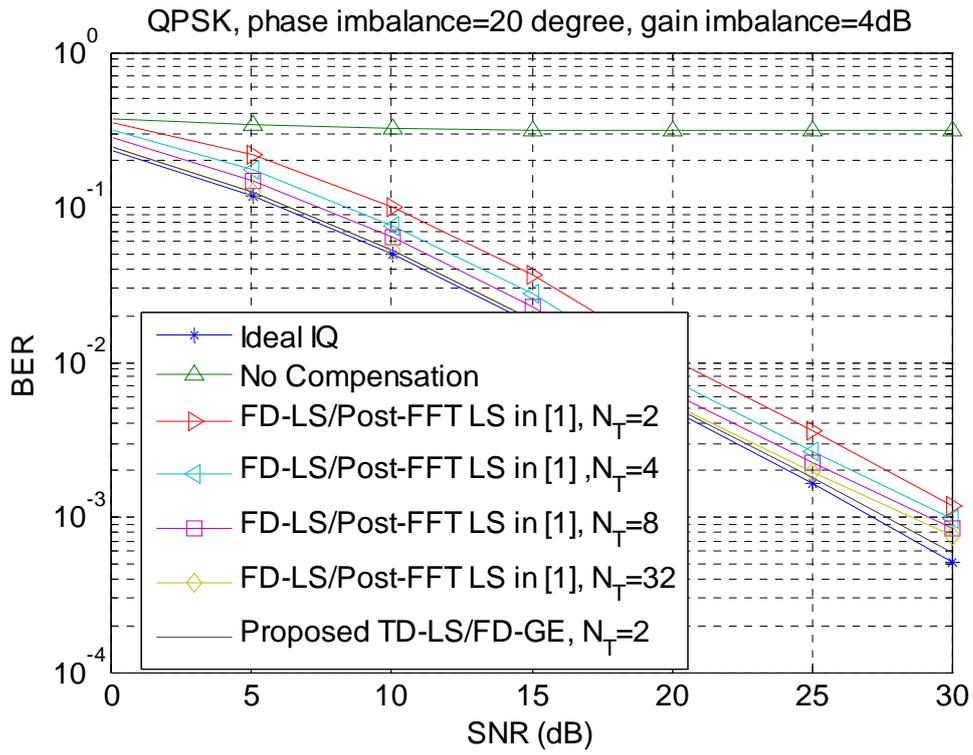

Fig. 4 Comparison of BER performance for the proposed scheme with two training OFDM symbols and the LS in [1] with different numbers of training symbols in the case of $\theta = 20^{\mathrm{o}}$ and $\alpha = 4$ dB.

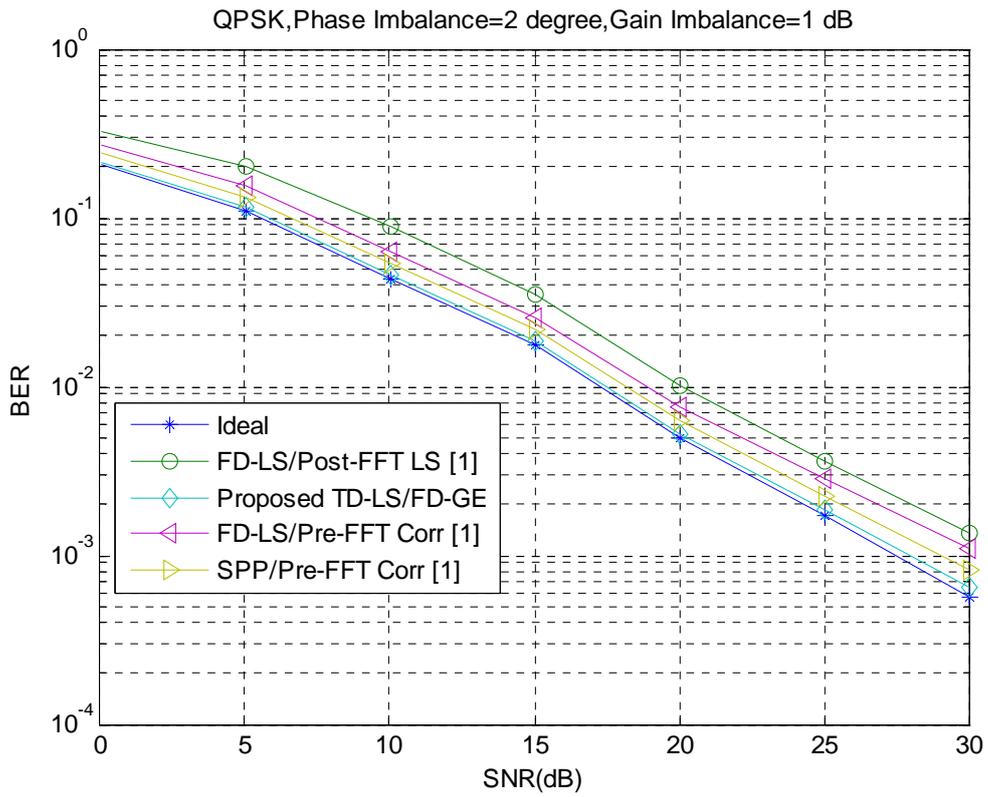

Fig. 5 Comparison of BER performance for the proposed scheme and three schemes in [1] for two

OFDM training symbols in the case of $\theta=2^o$ and $\alpha=1$dB.

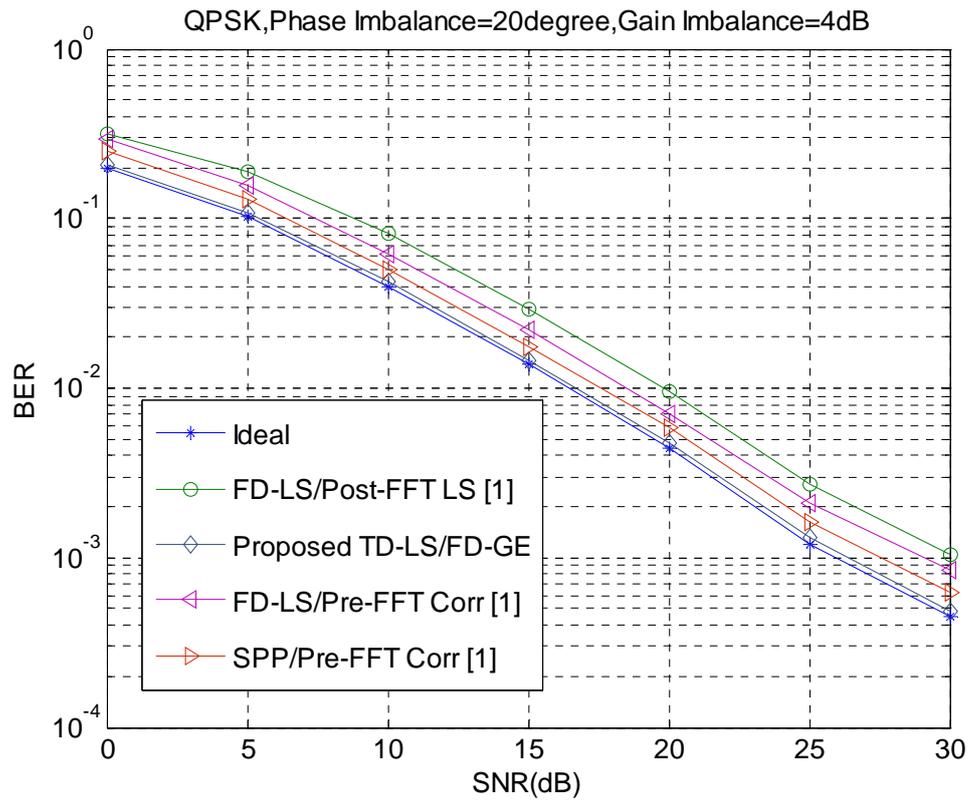

Fig. 6 Comparison of BER performance for the proposed scheme and three schemes in [1] for two OFDM training symbols in the case of $\theta = 20^{o}$ and $\alpha = 4$ dB.